\documentclass[journal]{IEEEtran}
\usepackage{graphicx}

\begin{document}

\title{Level Crossing Rate and Average Fade Duration \\ of the Double Nakagami-$m$ Random Process \\ and Application in MIMO
Keyhole Fading Channels}

\author{Nikola Zlatanov,~\IEEEmembership{}Zoran~Hadzi-Velkov,~\IEEEmembership{}and George K. Karagiannidis,~\IEEEmembership{\vspace{-0.7cm}}%
\thanks{Accepted for IEEE CommLetters}
\thanks{N. Zlatanov and Z. Hadzi-Velkov are with the Faculty of Electrical Engineering and Information Technologies, Ss. Cyril and Methodius University, Skopje, Email: zoranhv@feit.ukim.edu.mk, nzlatanov@manu.edu.mk}
\thanks{G. K. Karagiannidis is with the Department of Electrical and Computer Engineering, Aristotle University of Thessaloniki, Thessaloniki, Email: geokarag@auth.gr}
}

\markboth{ }{Shell \MakeLowercase{\textit{et al.}}: Bare Demo of
IEEEtran.cls for Journals} \maketitle

\begin{abstract}
We present novel exact expressions and accurate closed-form
approximations for the level crossing rate (LCR) and the average
fade duration (AFD) of the double Nakagami-$m$ random process.
These results are then used to study the second order statistics
of multiple input multiple output (MIMO) keyhole fading channels
with space-time block coding. Numerical and computer simulation
examples validate the accuracy of the presented mathematical
analysis and show the tightness of the proposed approximations.
\end{abstract}


\section{Introduction}
Recently, special attention has been given to the so-called
``multiplicative" fading models. The double Rayleigh (i.e.,
Rayleigh*Rayleigh) channel fading model has been found to be
suitable when both transmitter and receiver are moving \cite{x1}.
Moreover, it has also been recently used for keyhole channel
modeling of multiple-input multiple-output (MIMO) systems
\cite{x3}-\cite{x4}. Its extension, the double Nakagami-$m$ (i.e.,
Nakagami-$m$*Nakagami-$m$) fading model, has been considered in
\cite{x5}, where the fading between each pair of transmit and
receive antennas in presence of the ``keyhole" is characterized as
Nakagami-$m$ fading. However, all the above works describe and
utilize only the f\-irst order statistical properties of these
``multiplicative" fading models, such as the outage and the error
probabilities. But, knowledge of the second order statistics for
above fading models are equally important, and are applicable, for
example, in modeling and design of the multihop communications
systems \cite{x6}.

In this letter, we focus on the second order statistics of the
double Nakagami-$m$ random process, for which we determine exact
and approximate analytical solutions for its level crossing rate
(LCR) and average fade duration (AFD). Then we apply these results
to study the second order statistics of the keyhole channels
applicable to MIMO systems with space-time block coding (STBC),
operating in specif\-ic rich-scattering environments. Note that
although this work assumes independence among the channels,
similar analysis can be used to derive LCR and AFD in correlated
keyhole channels \cite{10a}.

\section{On the second order statistics of the double Nakagami-$m$ random process}
Let the double Nakagami-$m$ random process be def\-ined as
\begin{equation}\label{0}
Z(t)=X(t)Y(t) \,,
\end{equation}
where $X(t)$ and $Y(t)$ are a pair of independent Nakagami-$m$
distributed RVs with probability distribution functions (PDFs)
\begin{equation}\label{1}
f_{X}(x)=\left(\frac{m_X}{\Omega_X}\right)^{m_X}\frac{2 x^{2 m_X-1}}{\Gamma(m_X)}\exp\left(-\frac{m_X x^2}{\Omega_X}\right)
\end{equation}
and
\begin{equation}\label{2}
f_{Y}(y)=\left(\frac{m_Y}{\Omega_Y}\right)^{m_Y}\frac{2 y^{2 m_Y-1}}{\Gamma(m_Y)}\exp\left(-\frac{m_Y y^2}{\Omega_Y}\right) \,,
\end{equation}
where $\Omega_X = E[X^2]$, $\Omega_Y = E[Y^2]$, and $m_X$ and
$m_Y$ are the fading severity parameters, where $E[\cdot]$ means
expectation.

If $X(t)$ and $Y(t)$ are signal envelopes in some scattering radio
channel exposed to the Doppler effect due to stations' relative
mobility, then $X(t)$ and $Y(t)$ are time-correlated random
processes. Considering a f\-ixed-to-mobile channel, each scattered
component of $X(t)$ and $Y(t)$ has some resulting Doppler spectra
with maximum Doppler frequency shift $f_{mx}$ and $f_{my}$,
respectively. It was shown in \cite{7a} that, under such
conditions, the envelopes time derivatives $\dot X$ and $\dot Y$
are independent from their respective envelopes, while following
zero-mean Gaussian PDFs with respective variances
\begin{eqnarray}\label{3}
\sigma_{\dot X}^2=(\pi f_{mx})^2\Omega_X/m_X, \quad \sigma_{\dot
Y}^2=(\pi f_{my})^2\Omega_Y/m_Y \,.
\end{eqnarray}

\subsection{Second order statistics}
The LCR of $Z$ at threshold $z$ is def\-ined as the rate at which
the random process crosses level $z$ in the negative direction. To
extract LCR, we need to determine the joint PDF of $Z$ and $\dot
Z$, $f_{Z \dot Z }(z,\dot z)$, and apply the Rice's formula
\begin{equation}\label{14}
N_Z(z)=\int_0^\infty \dot z f_{Z \dot Z}(\dot z,z) d\dot z \,.
\end{equation}
The above expression can be rewritten as
\begin{equation}\label{15}
N_Z(z)=\int_0^\infty\left(\int_0^\infty\dot z f_{\dot Z|Z X}(\dot z|z,x)d\dot z\right) f_{Z|X}(z|x)f_X(x)dx
\end{equation}
where $f_{\dot Z | ZX}(\cdot,\cdot,\cdot)$ is the conditional PDF
of $\dot Z$ conditioned on $Z$ and $X$. This conditional PDF can
be determined by f\-inding the time derivative of both sides of
(\ref{0}),
\begin{eqnarray}\label{16}
\dot{Z}=Y\dot X+X\dot Y=\frac{Z}{X}\dot X+ X\dot Y \,,
\end{eqnarray}
from which it is easily seen that, for f\-ixed $Z=z$ and $X=x$,
the time derivative $\dot Z$ is a zero-mean Gaussian RV with
variance $\sigma_{\dot Z|Z X}^2=z^2\sigma_{\dot X}^2/x^2 +
x^2\sigma_{\dot Y}^2$. Now, the bracketed integral in (\ref{15})
can be solved as
\begin{equation}\label{18}
\int_0^\infty\dot z f_{\dot Z|ZX}(\dot z|z,x)d\dot
z=\frac{\sigma_{\dot Z|ZX}}{\sqrt{2\pi}} \,.
\end{equation}
The conditional PDF of $Z$ for some f\-ixed $X = x$, $f_{Z |
X}(z|x)$, is determined by simple transformation of RVs,
$f_{Z|X}(z|x)=f_Y(z/x)/x$. Substituting (\ref{18}) into
(\ref{15}), after some algebraic manipulations, we obtain the
exact solution for the LCR
\begin{eqnarray}\label{20}
N_Z(z)=\frac{1}{\sqrt{2\pi}}\frac{4 z^{2 m_Y-1} \; \sigma_{\dot Y}
}{\Gamma(m_X)\Gamma(m_Y)}\left(\frac{m_X}{\Omega_X}\right)^{m_X}
\left(\frac{m_Y}{\Omega_Y}\right)^{m_Y} \quad \nonumber\\
\times \int_0^\infty \sqrt{1+\frac{z^2}{x^4}
\left(\frac{\sigma_{\dot X}}{\sigma_{\dot
Y}}\right)^2}x^{2(m_X-m_Y)} \; e^{-\big(\frac{m_X x^2}{\Omega_X} +
\frac{m_Y z^2}{\Omega_Y x^2}\big)} dx
\end{eqnarray}
The above integral can be evaluated numerically with desired
accuracy (e.g. by using some common software such as Mathematica).
Alternatively, one can apply the Laplace approximation to obtain a
highly accurate closed-form solution of (\ref{20}) - as presented
in the following subsection.

The AFD of $Z$ at threshold $z$ is def\-ined as the average time
that the double Nakagami-$m$ random process remains below level
$z$ after crossing that level in the downward direction,
\begin{equation}\label{20a}
T_Z(z)=\frac{F_Z(z)}{N_Z(z)} \,,
\end{equation}
where $F_Z(z)$ denotes the CDF
of $Z$, which was derived only recently in closed-form for
$N$*Nakagami random process \cite{8a}. For the double Nakagami
random process, it attains the form
\begin{equation}\label{20b}
F_Z(z)=\frac{1}{\Gamma(m_X)\Gamma(m_Y)} \, G_{1,3}^{2,1} \left[z^2\frac{m_X m_Y}{\Omega_X\Omega_Y}\Bigg |
\begin{array}{cc}
 \qquad 1 \\
m_X,m_Y, 0
\end{array}
 \right],
\end{equation}
where $\Gamma(\cdot)$ and $G[\cdot]$ are gamma and Meijer's $G$ functions. \vspace{-0.2cm}
\subsection{Laplace approximation}
Using \cite{9a}, the Laplace type integral can be approximated as
\begin{equation}\label{21}
\int_0^\infty g(x) \; e^{-\lambda
f(x)}dx\approx\sqrt{\frac{2\pi}{\lambda}}\frac{g(x_0)}{\sqrt{f^{''}(x_0)}}
\, e^{-\lambda f(x_0)} \,,
\end{equation}
when the real valued parameter $\lambda$ is very large (i.e.,
$\lambda\to \infty$). In (\ref{21}), $f(x)$ and $g(x)$ are
real-valued functions of $x$ and $x_0$ is the point at which
$f(x)$ has an absolute minimum (known as the interior critical
point of $f(x)$). Note, that $f''(x)$ denotes the second
derivative of $f(x)$ with respect to $x$. It was observed that
above approximation is very accurate even for small values of
$\lambda$ \cite{9a}. Comparing (\ref{21}) and (\ref{20}), these functions
are set as
\begin{equation}\label{22}
f(x)=\frac{m_X
x^2}{\Omega_X}+\frac{m_Y}{\Omega_Y}\left(\frac{z}{x}\right)^2-\ln(x^{2(m_X-m_Y)}) \,,
\end{equation}
\begin{equation}\label{22a}
g(x)=\sqrt{1+\frac{z^2}{x^4}\left(\frac{\sigma_{\dot
X}}{\sigma_{\dot Y}}\right)^2} \,,
\end{equation}
whereas the second derivative of the former is $f^{''}(x)= 2
m_X/\Omega_X + 6 (m_Y z^2)/(\Omega_Y x^4) + 2 (m_X-m_Y)/x^2$ and
$\lambda = 1$. The  critical point of $f(x)$ is determining as the
value of $x$ for which $\partial f/\partial x = 0$, i.e.,
\begin{eqnarray}\label{24}
x_0 = \Big [ \frac{1}{2 m_X \Omega_Y} \Big(
\Omega_X\Omega_Y(m_X-m_Y)
\qquad \qquad \qquad \quad \nonumber \\
+ \sqrt{\Omega_X^2\Omega_Y^2(m_X-m_Y)^2+4m_X m_Y\Omega_X\Omega_Y
z^2} \,  \Big ) \Big ]^\frac12 .
\end{eqnarray}
Using (\ref{22})-(\ref{24}), the approximate closed-form solutions
for the LCR and the AFD are respectively obtained as
\begin{eqnarray}\label{28}
N_Z(z) \approx \frac{4 z^{2 m_Y-1} \; \sigma_{\dot Y}}{\Gamma(m_X)\Gamma(m_Y)} \qquad \qquad \qquad \qquad \qquad \qquad \quad \nonumber\\
\times
\left(\frac{m_X}{\Omega_X}\right)^{m_X}\left(\frac{m_Y}{\Omega_Y}\right)^{m_Y}
\frac{g(x_0)}{\sqrt{f^{''}(x_0)}} \, e^{-f(x_0)} \,, \quad
\end{eqnarray}
\vspace{-0.25cm}
\begin{eqnarray}\label{28a}
T_Z(z) \approx \frac{1}{4 z^{2 m_Y-1} \; \sigma_{\dot Y}}
\left(\frac{\Omega_X}{m_X}\right)^{m_X}
\left(\frac{\Omega_Y}{m_Y}\right)^{m_Y} \qquad \qquad
\nonumber\\
\times \frac{\sqrt{f^{''}(x_0)} \;\; e^{f(x_0)}}{g(x_0)} \;\;
G_{1,3}^{2,1} \left[z^2 \frac {m_X m_Y}{\Omega_X \Omega_Y}\Bigg |
\begin{array}{cc}
 \qquad 1 \\
m_X,\,m_Y, \,0
\end{array}
 \right] .
\end{eqnarray}
Although substitution of $f(x_0)$, $f^{''}(x_0)$ and $g(x_0)$ into
(\ref{28}) and (\ref{28a}) is omitted for brevity, we emphasize
that the threshold $z$ appears only as the ratio
$z^2/((\Omega_X/m_X)(\Omega_Y/m_Y))$. \vspace{-0.2cm}
\section{MIMO STBC Communication over Keyhole Fading Channels}
Potentials of MIMO communications systems are not always
achievable even for a fully uncorrelated transmit and receive
channels, which is attributed to the rank def\-iciency of the MIMO
channels known as the keyhole or pinhole effect \cite{x3}. The
existence of the keyhole MIMO channels has been proposed and
demonstrated through physical examples, where, although spatially
uncorrelated, these channels still have a single degree of freedom
\cite{x3}-\cite{x4}. Under the keyhole effect, the entries of the
channel matrix, $ \mathbf H$, follow statistics described as a
product of two independent single-path gains. \vspace{-0.2cm}
\subsection{The MIMO keyhole channel model}
From \cite{x5}, the complex path gain of baseband equivalent
signal transmitted over the channel between the  $i$-th transmit
and the $j$-th receive antenna at arbitrary moment $t$ is
expressed as $1\leq i\leq M , 1\leq j\leq N$
\begin{equation}\label{00}
h_{ij}(t)=\alpha_i(t)\beta_j(t)e^{j(\phi_i(t)+\psi_j(t))} ,
\end{equation}
where $\left\{\alpha_i(t) e^{j\phi_i(t)}\right\}_{i=1}^M$  are the
complex path gains introduced by the rich-scattered channel from
the $i$-th transmitting antenna to the ``keyhole", and
$\left\{\beta_j(t) e^{j\psi_j(t)}\right\}_{j=1}^N$ are the complex
path gains introduced by the rich-scattered channel from the
``keyhole" to the $j$-th receiving antenna. Phases
$\left\{\phi_i(t)\right\}_{i=1}^M$ and
$\left\{\psi_j(t)\right\}_{j=1}^N$ are independent and uniformly
distributed over $[0,2\pi)$. The amplitudes $\left\{\alpha_i(t)
\right\}_{i=1}^M$ and $\left\{\beta_j(t) \right\}_{j=1}^N$ are
i.i.d. Nakagami-$m$ RVs. The fading severity parameters of
$\alpha_i(t)$ are equal to $m_T$, whereas $\Omega_T =
E[\alpha_i^2]$ for all $i$. Similarly, the fading severity
parameters of $\beta_j(t)$ are equal to $m_R$, whereas $\Omega_R =
E[\beta_j^2]$ for all $j$. Assuming mobility of both the
transmitter and the receiver with respect to the ``keyhole", all
channel gains are time-correlated random processes with maximum
Doppler shifts $f_{\alpha_i}=f_{\alpha}$  and
$f_{\beta_i}=f_{\beta}$, respectively. Under such conditions, the
time derivatives $\dot\alpha_i$ and $\dot\beta_j$ are independent
from $\alpha_i$ and $\beta_j$, respectively, and both follow
zero-mean Gaussian PDFs with variances given by (\ref{3}),
$\sigma_{\dot \alpha_i}^2=(\pi f_\alpha)^2\Omega_T/m_T$, $1 \leq i
\leq M$ and $\sigma_{\dot \beta_j}^2=(\pi f_\beta)^2\Omega_R/m_R$,
$1 \leq j \leq N$.

\begin{figure}
\centering
\includegraphics[width=3.1in]{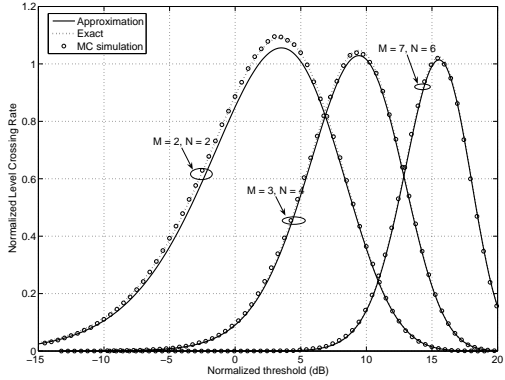}
\vspace{-0.5cm} \caption{Normalized LCR for various number of
transmit and receive antennas} \label{fig_4} \vspace{-0.4cm}
\end{figure}
\vspace{-0.35cm}
\subsection{Orthogonal space-time block coding and decoding }
The orthogonal space–time block encoding and decoding (signal
combining) transform a MIMO fading channel into an equivalent
single-input-single-output (SISO) fading channel with a path gain
of the squared Frobenius norm of the MIMO channel matrix $\mathbf
H(t)=\left[h_{ij}(t)\right]_{M\times N}$ \cite{x5},
\begin{equation}
||\mathbf H(t)||_F^2=\sum_{i=1}^M\sum_{j=1}^N |h_{ij}(t)|^2  =\left(\sum_{i=1}^M\alpha_i^2(t)\right)\left(\sum_{j=1}^N\beta_j^2(t)\right)
\end{equation}
at arbitrary moment $t$. After space-time block decoding, the
instantaneous output signal-to-noise ratio (SNR) per symbol is
given by
\begin{equation}\label{6}
\gamma(t)=\frac{\bar\gamma}{M R}||\mathbf H(t)||_F^2 \,,
\end{equation}
where $\bar\gamma  = E_s/N_0$ is the average SNR per receive
antenna, and $R$ is the rate of the STBC. \vspace{-0.25cm}
\subsection{Second order statistics of output SNR}
We introduce the auxiliary random process $Z(t)$ def\-ined by
\begin{equation}\label{7}
Z(t)=\sqrt{||\mathbf H(t)||_F^2}=X(t)Y(t) \,,
\end{equation}
where $X(t) = \sqrt{\sum_{i=1}^M\alpha_i^2(t)}$ and
$Y(t)=\sqrt{\sum_{j=1}^N\beta_j^2(t)}$ are again Nakagami-$m$
distributed with  PDFs given by (\ref{1}) and (\ref{2}),
respectively, with  $m_X = M m_T$, $\Omega_X = M \Omega_T$, $m_Y =
N m_R$ and $\Omega_Y = N \Omega_R$. The time derivatives $\dot X$
and $\dot Y$ are independent from $X$ and $Y$, respectively, and
both follow the zero-mean Gaussian PDF with variances given by
(\ref{3}), $\sigma_{\dot X}^2=\sigma_{\dot \alpha_i}^2=(\pi
f_\alpha)^2\Omega_T/m_T$ and $\sigma_{\dot Y}^2=\sigma_{\dot
\beta_j}^2=(\pi f_\beta)^2\Omega_R/m_R$.

Hence, the random process $Z(t)$, def\-ined by (\ref{7}), is a
double Nakagami-$m$ process for which we can apply the analytical
framework of Section II  to determine its exact and approximate
LCR and AFD   by using (\ref{20}), (\ref{20b}), (\ref{28}) and
(\ref{28a}). With above in mind, the LCR and the
AOD\footnote{Instead of the term "average fade duration (AFD)",
the term "average outage duration (AOD)" is used here.} of
instantaneous output SNR, given by (\ref{6}), are respectively
determined as
\begin{eqnarray}\label{31}
N_\gamma(\gamma)&=&N_Z (\sqrt {\gamma M R /\bar\gamma}) \,, \\
T_\gamma(\gamma)&=&T_Z (\sqrt {\gamma M R /\bar\gamma}) \,.
\end{eqnarray}

\vspace{-0.2cm}
\section{Numerical Results}
We present several numerical examples for the LCR
and the AFD of the STBC MIMO communications system operating over
a keyhole fading channel. The mobile transmitter and the mobile
receiver are assumed to introduce same maximum Doppler shifts due
to same relative speeds with respect to the ``keyhole", yielding
$f_{\alpha}=f_{\beta}=f_m$.

Figs. 1 and 2 depict the normalized LCR ($N_\gamma/f_m$) and
normalized AFD ($T_\gamma \, f_m$) of the instantaneous output SNR
vs. normalized SNR threshold. The normalized SNR threshold
($x$-axis) is calculated as $10\log[\gamma \, M R /(\bar\gamma
(\Omega_T/m_T)(\Omega_R/m_R))]$. The results are obtained for
three different pairs of number of transmit and receive antennas
$(M,N)$, appearing as curve parameters. For each pair $(M,N)$, the
three comparative curves on both f\-igures indicate excellent
match between the exact and the approximate solutions for the two
statistical parameters, both of which are validated by Monte Carlo
simulations. \vspace{-0.1cm}

\begin{figure}
\centering
\includegraphics[width=3.1in]{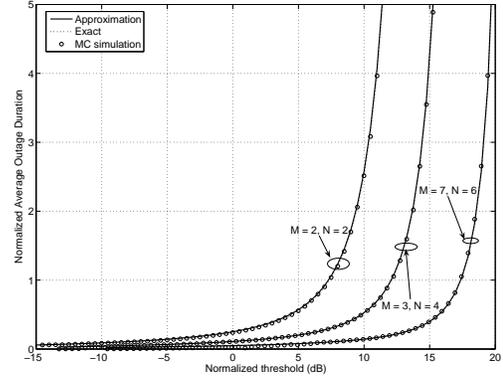}
\vspace{-0.5cm} \caption{Normalized AOD for various number of
transmit and receive antennas} \label{fig_5} \vspace{-0.4cm}
\end{figure}


\end{document}